\documentclass[11pt,twoside]{article}
\usepackage{CAGN2016}
\usepackage{graphicx}

\usepackage[T1]{fontenc} 

\usepackage{latexsym}
\usepackage{verbatim}

\begin{document}

\vskip 1.0cm
\markboth{G.~Dom\'inguez-Guzm\'an et al.}{The abundances in H~{\sc ii} regions of the Magellanic Clouds}
\pagestyle{myheadings}
%
%
\vspace*{0.5cm}
\parindent 0pt{Contributed  Paper}


\vspace*{0.5cm}
\title{The abundances of O, N, S, Cl, Ne, Ar, and Fe in H~{\sc ii} regions of the Magellanic Clouds}

\author{G.~Dom\'inguez-Guzm\'an$^1$, M.~Rodr\'iguez$^1$, C. Esteban$^{2,3}$ and J.~Garc\'ia-Rojas$^{2,3}$}
\affil{
$^{1}$Instituto Nacional de Astrof\'isica, \'Optica y Electr\'onica,  Apdo. Postal 51 y 216, Puebla, Mexico \\
$^{2}$Instituto de Astrof\'isica de Canarias, E-38200, La Laguna, Tenerife, Spain \\
$^{3}$Departamento de Astrof\'isica, Universidad de La Laguna, E-38206, La Laguna, Tenerife, Spain}

\begin{abstract}
We use very deep spectra obtained with the Ultraviolet-Visual Echelle Spectrograph in the Very Large Telescope in order to determine the physical conditions, the chemical abundances and depletion factors of four H~{\sc ii} regions of the Large Magellanic Cloud and four H~{\sc ii} regions of the Small Magellanic Cloud. The spectral range covered is 3100-10400 $\mathring{\mbox{A}}$ with a resolution of $\Delta\lambda \sim \lambda/8800$. We measure the intensity of up to 200 emission lines in each object. Electron temperature and electron density are determined using different line intensity ratios. The ionic and total abundances are derived using collisionally excited lines for O, N, S, Cl, Ne, Ar, and Fe. The uncertainties for all the computations are calculated using Monte Carlo simulations. This is the largest available set of high quality spectra for H~{\sc ii} regions in the Magellanic Clouds. Thus, we can derive chemical abundances and depletion factors and constrain their variations across each galaxy with better accuracy than previous studies. In particular, we find that the amount of Fe depleted on to dust grains in the H~{\sc ii} regions of the Magellanic Clouds is similar to that found in Galactic H~{\sc ii} regions.
\end{abstract}

\section{Introduction}
The Magellanic Clouds (MCs) are the laboratories for studies in extragalactic H~{\sc ii} regions at lower metallicities than in the Milky Way. Deep optical spectra of H~{\sc ii} regions in these galaxies would allow us to detect and measure relatively faint iron lines and thus, study the properties of dust in ionized nebulae through the analysis of the iron depletion factor. This factor is defined as the ratio between the expected abundance of iron and the one measured in the gas phase. The study of the dust in ionized nebulae of the MCs will provide clues on the processes responsible for the formation and evolution of the grains and the role played by metallicity in the efficiency of these processes. Rodr\'iguez \& Rubin (2005) and Delgado-Inglada et~al. (2011) found that ionized nebulae with lower metallicities show lower depletions factors. We extend the sample for H~{\sc ii} regions in the MCs in order to better understand  this behavior.
\section{Observational data}
\label{data}

\begin{figure}  
\begin{center}
\hspace{0.25cm}
\includegraphics[height=8cm]{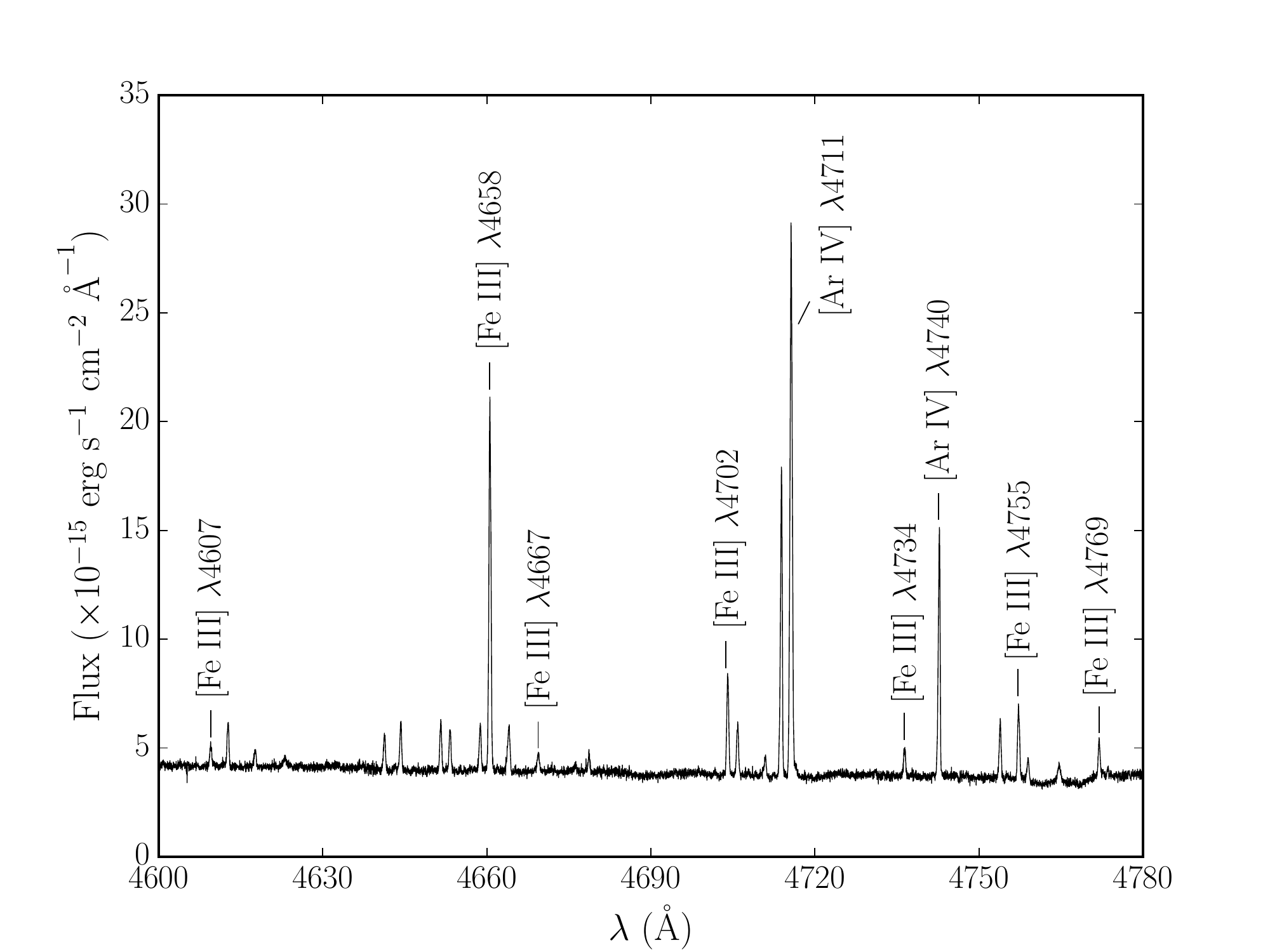}
\caption{Part of the deep spectrum of the H{\sc ii} N81 in the Small Magellanic Cloud.}
\label{N81spec_fig1}
\end{center}
\end{figure}

We obtained echelle spectra of 8 H~{\sc ii} regions in the MCs, 4 in the Small Magellanic Cloud (SMC)  and 4 in the Large Magellanic Cloud (LMC). The data were taken with the Ultraviolet-Visual Echelle Spectrograph (UVES) in the Very Large Telescope (VLT) in Chile. The spectral range covered was from 3100 $\mathring{\mbox{A}}$ to 10400 $\mathring{\mbox{A}}$ with a spectral resolution of $\Delta\lambda \sim \lambda/8800$. The atmospheric dispersion corrector was used to keep the same observed region within the slit since the MCs are observed at relatively high airmass (between 1.35 and 1.88). The slit position for each object was chosen to have enough spectral resolution and high signal-to-noise.\\
\\
The data reduction was performed using the public UVES pipeline under the {\sc gasgano} graphic user interface through the standard procedure of bias and aperture subtraction, flat fielding and wavelength calibration. For flux calibration we used the available tasks in the {\sc iraf}\footnote{{\sc iraf} is distributed by the National Optical Astronomy Observatories, which are operated by the Association of Universities for Research in Astronomy, Inc., under cooperative agreement with the National Science Foundation.} software package using the standard stars HR718, HR 3454 and HR9087. Figure~\ref{N81spec_fig1} shows part of the spectrum of N81 where we identify several iron emission lines. Line intensities were measured by integrating the flux above the continuum defined by two points on each side of the emission lines. Flux uncertainties were computed by adding quadratically line fluxes and flux calibration uncertainties. We finally include the uncertainty associated to the propagation in the reddening coefficient.


\section{Results}
\label{results}

The calculations of physical conditions and ionic and total abundances were carried out with {\sc PyNeb} (Luridiana et al. 2015). The uncertainties were obtained through Monte Carlo simulations. We determined four diagnostics for electron density using 
sulfur, oxygen, chlorine and argon lines. We decided to use the average electron density only from $n_{\mbox{\small{e}}}$([O~{\sc ii}]) and $n_{\mbox{\small{e}}}$([S~{\sc ii}]) diagnostics since $n_{\mbox{\small{e}}}$([Cl~{\sc iii}]) and $n_{\mbox{\small{e}}}$([Ar~{\sc iv}]) give higher densities owing to they are less sensitive at lower densities. We computed electron temperature from five different diagnostics: $T_{\mbox{\small{e}}}$([N~{\sc ii}]), $T_{\mbox{\small{e}}}$([O~{\sc ii}]), $T_{\mbox{\small{e}}}$([O~{\sc iii}]), $T_{\mbox{\small{e}}}$([S~{\sc iii}]) and $T_{\mbox{\small{e}}}$([Ar~{\sc iii}]). We decided to use  $T_{\mbox{\small{e}}}$([N~{\sc ii}]) as the representative temperature for the low ionization degree zone because the emission lines used to determine $T_{\mbox{\small{e}}}$([O~{\sc ii}]) are affected by recombination and density variations. For high ionization degree zone we decided to use $T_{\mbox{\small{e}}}$([O~{\sc iii}]) because some of the emission lines used to determine $T_{\mbox{\small{e}}}$([S~{\sc iii}]) are affected by telluric absorption and $T_{\mbox{\small{e}}}$([Ar~{\sc iii}]) gives higher uncertainties than $T_{\mbox{\small{e}}}$([O~{\sc iii}]).\\

Once we determined the physical conditions, we computed the ionic abundances for all the available ions assuming the two-zone scheme. For oxygen we added directly O$^+$/H$^+$ and O$^{++}$/H$^+$ when He II lines are not observed; otherwise we used the ionization corrector factor (ICF) given by Delgado-Inglada et al. (2014). For nitrogen we used the classical ICF, N/O = N$^+$/O$^+$, as suggested by Delgado-Inglada et al. (2015). For iron we used the ICFs given by Rodr\'iguez \& Rubin (2005). We used the two ICFs they recommend because they give us extreme values of the total iron abundance and can be used to constrain the true values of the Fe abundances in the gas. For the rest of elements we used the ICFs given by Delgado-Inglada et al. (2014). The uncertainties due the ICFs are not considered.\\

\begin{figure}
\begin{center}
\hspace{0.25cm}
\includegraphics[height=8cm]{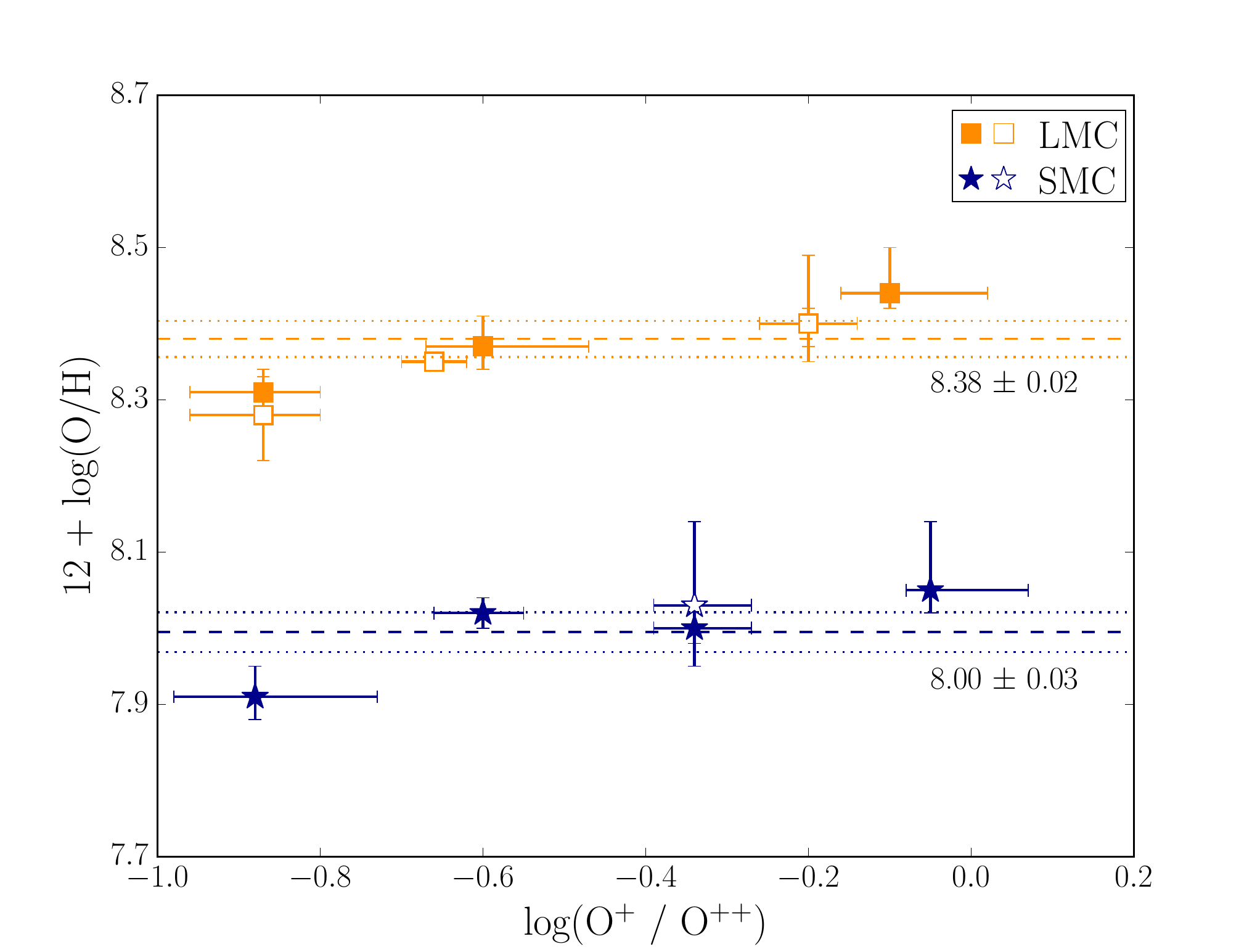}
\caption{Total oxygen abundance as a function of the ionization degree. The orange squares symbols are H~{\sc ii} regions from the LMC and the blue stars symbols are H~{\sc ii} regions from the SMC. The filled symbols represent our main sample and the empty ones are objects collected from the literature. The orange and blue dashed lines correspond to the average of the total oxygen abundance for the LMC and the SMC, respectively.}
\label{Oxab_fig2}
\end{center}
\end{figure}

\begin{figure}  
\begin{center}
\hspace{0.25cm}
\includegraphics[height=8.4cm]{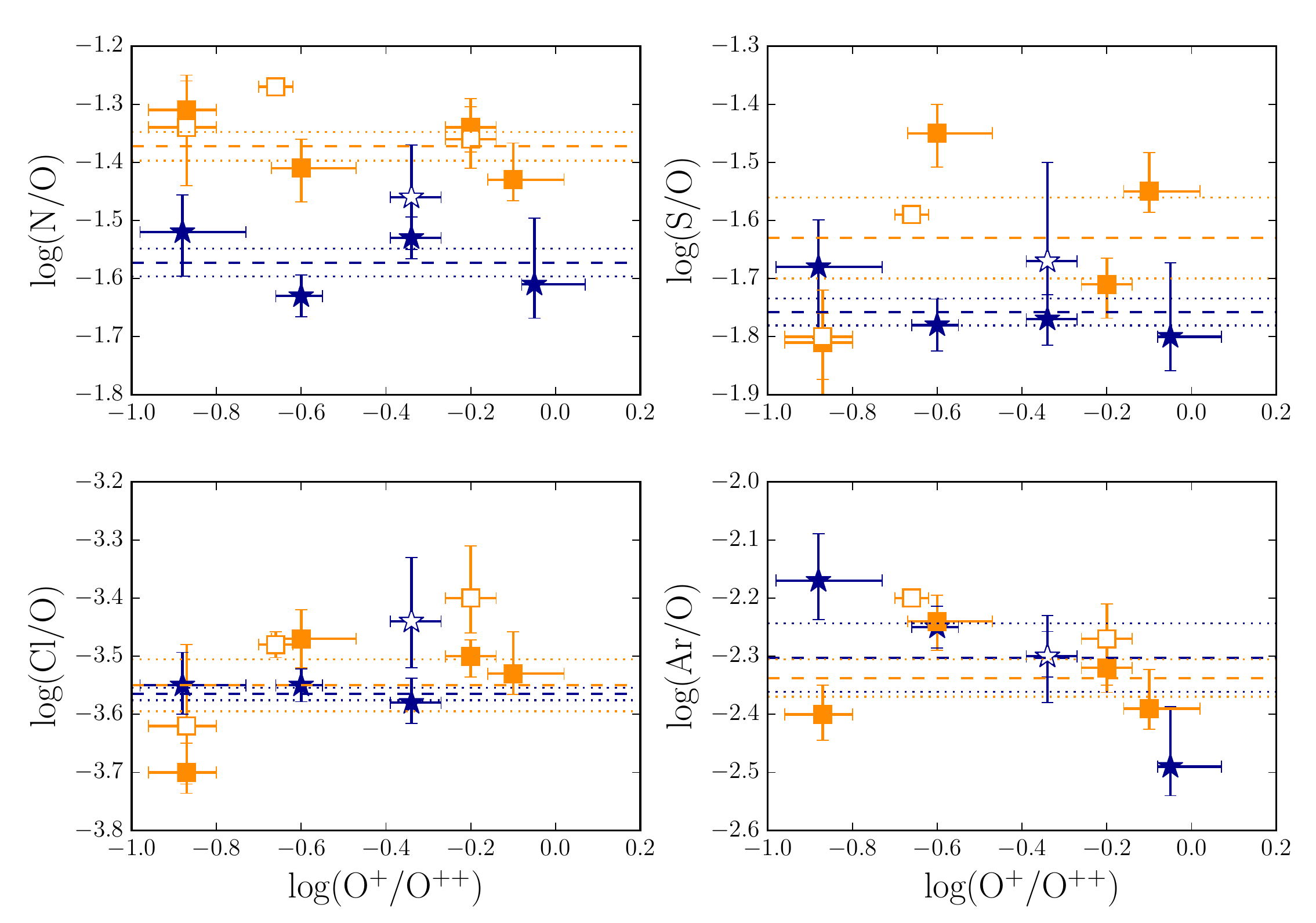}
\caption{Total abundances of N, S, Cl, Ar as a function of the ionization degree. The color code is the same as in Figure~\ref{Oxab_fig2}. }
\label{abund_fig3}
\end{center}
\end{figure}

Figure~\ref{Oxab_fig2} shows the oxygen abundance as a function of the ionization degree for all the objects (filled symbols). The squares are H~{\sc ii} regions from the LMC and the stars are H~{\sc ii} regions from the SMC. We include other objects from the literature as empty symbols (Tsamis et al. 2003, Naz\'e et al. 2003 and Peimbert 2003). We can see that the O abundance is higher for the LMC than the SMC by $\sim$ 0.38 dex. In Figure~\ref{abund_fig3} we plot the total abundance for the other elements with respect to oxygen as a function of the ionization degree. The S and Cl abundances are similar in both clouds within the uncertainties; the same behavior is seen for the Ne abundances. For N abundances it seems to be a separation between the two MCs due to differences in the chemical evolution of each galaxy. For Ar abundances there is a problem with the ICF since there is a trend with the ionization degree.\\

\begin{figure}  
\begin{center}
\includegraphics[angle=0,height=8.cm]{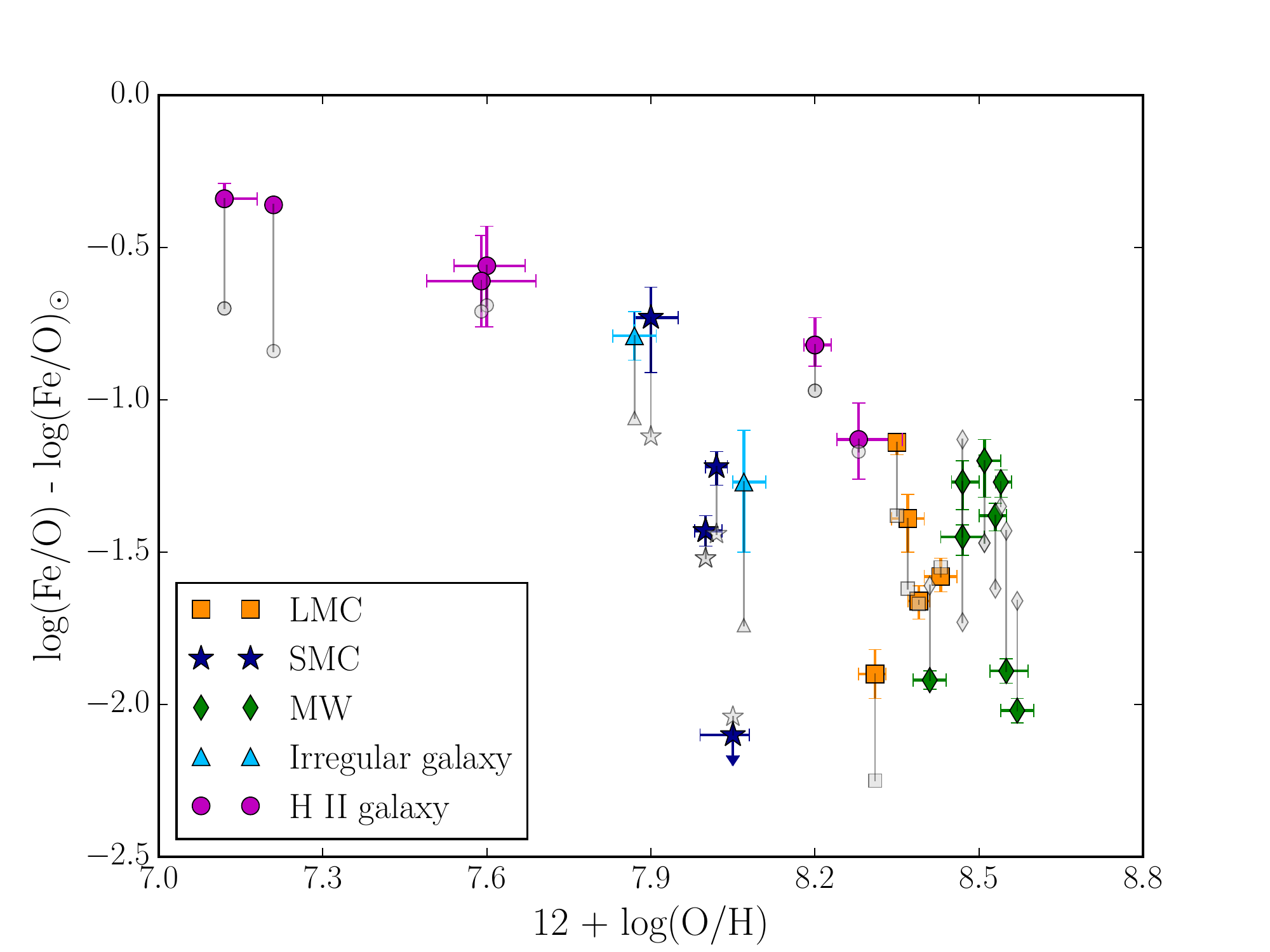}
\caption{ The depletion factor for iron as a function of the oxygen abundance. The colored symbols are for the ICF from equation (2) of  Rodr\'iguez \& Rubin (2005) and the gray ones are the ICF from equation (3). }
\label{DF_fig4}
\end{center}
\end{figure}

Figure~\ref{DF_fig4} shows the values of the the depletion factor for Fe derived with the two ICFs as a function of the oxygen abundance for all the objects. The colored symbols are for the ICF from equation (2) and the gray symbols are for the ICF from equation (3) of Rodriguez \& Rubin (2005). We include Galactic H~{\sc ii} regions (green diamonds), H~{\sc ii} galaxies (purple circles), irregular galaxies (cyan triangles). We performed the analysis in the same way than for the main sample. We see that the iron depletions into dust grains in H~{\sc ii} regions of the LMC are similar to those found in Galactic nebulae in previous works.  The SMC also shows this behavior but in a wider range of iron depletions. Besides, there is an object: N88A, that follows the trend that H~{\sc ii} with lower metallicities show lower depletions factors.

\section{Conclusions}
\label{discussion}
We present new determinations of chemical abundances of 8 H~{\sc ii} regions in the MCs, 4 in the LMC and 4 in the SMC, using deep echelle spectra taken at the VLT. The main conclusions are:
\begin{quote}
$\bullet$ O/H is higher in the LMC by $\sim$0.4 dex.
\\ \\
$\bullet$ Cl/O and Ne/O are similar in both clouds within the uncertainties.
\\ \\
$\bullet$ S/O seems to be a separation between the two MCs.
\\ \\
$\bullet$ N/O is higher by $\sim$0.2 dex in the LMC.
\\ \\
$\bullet$ The amount of iron depletion into dust grains in H~{\sc ii} regions of the MCs is similar to that found in Galactic H~{\sc ii} regions.
\end{quote}

\acknowledgments We acknowledge support from Mexican CONACYT grant CB-2014-240562 and from MINECO under grant AYA2015-65205-P. \\
G.D.-G. acknowledges support from CONACYT grant 297932.

\end{document}